\documentclass[a4paper,11pt]{article}
\usepackage{pos}

\usepackage{multirow}
\usepackage{makecell}
\allowdisplaybreaks

\usepackage{color, colortbl,xcolor}

\title{On type IIB AdS$_{3}$ flux vacua with scale separation and integer conformal dimensions}

\author{\'Alvaro Arboleya}
\author*{Adolfo Guarino}
\author{Matteo Morittu}

\affiliation{Departamento de F\'isica, Universidad de Oviedo,\\
Avda. Federico Garc\'ia Lorca 18, 33007 Oviedo, Spain.}

\affiliation{Instituto Universitario de Ciencias y Tecnolog\'ias Espaciales de Asturias (ICTEA) \\
Calle de la Independencia 13, 33004 Oviedo, Spain.}

\emailAdd{arboleyaalvaro@uniovi.es}
\emailAdd{adolfo.guarino@uniovi.es}
\emailAdd{morittumatteo@uniovi.es}

\abstract{We review recent progress in constructing type IIB AdS$_{3}$ flux vacua that exhibit parametrically-controlled scale separation and come along with integer-valued conformal dimensions for the would-be dual CFT$_{2}$ operators. We comment on the anisotropic nature of the associated internal spaces, as well as on the existence of polynomial shift symmetries underlying the mass spectra. We conclude by outlining open questions and potential directions for future research, partially inspired by the Swampland program.}

\FullConference{Proceedings of the Corfu Summer Institute 2024 "School and Workshops on Elementary Particle Physics and Gravity" (CORFU2024)\
12 - 26 May, and 25 August - 27 September, 2024\\
Corfu, Greece\\}

\tableofcontents

\begin{document}
\maketitle

\section{Motivation and outlook}

One of the main challenges in string compactifications is to achieve moduli stabilisation in a vacuum that admits a parametrically-controlled scale separation between an external spacetime, typically anti de-Sitter (AdS), and the internal space of extra dimensions predicted by string theory. The most symmetric, perhaps better understood, compactifications involve spheres as internal spaces. A standard compactification of eleven-dimensional or type II supergravity on an internal sphere (and products thereof) typically gives rise to a lower-dimensional supergravity with the maximal amount of supersymmetry (\textit{i.e.} $32$ supercharges). These supergravities, also known as maximal supergravities, consist of just one multiplet, the supergravity multiplet, and come along with vector fields and a single parameter in the Lagrangian: the lower-dimensional gauge coupling $g$. Any AdS vacuum solution of the lower-dimensional supergravity will have a radius $\,L\,$ (inverse of the Hubble scale) of the form
\begin{equation}
L \propto \frac{1}{g} \ .
\end{equation}
Therefore, if one is interested in large AdS external spacetimes, one must take the limit $\,g\rightarrow 0\,$ which implies that the lower-dimensional maximal supergravity becomes \textit{ungauged}. In this limit, exact (exceptional) global symmetries emerge and the maximal supergravities are conjectured to be in the Swampland \cite{Montero:2022ghl}. Also, in a standard sphere compactification, the size of the internal sphere is also set by the lower-dimensional gauge coupling (recall that there is no other parameter in the theory). In particular,
\begin{equation}
L_{\textrm{sphere}} \sim \frac{1}{g} \quad\quad \Rightarrow \quad\quad L  \sim L_{\textrm{sphere}} \ ,
\end{equation}
and it is not possible to achieve scale separation between the (external) AdS spacetime and the (internal) sphere in the compactification.

A more promising avenue to achieve scale separation is to consider group manifolds (instead of spheres) as internal spaces for the compactification \cite{Scherk:1979zr,Hull:2005hk}. Group manifolds come along with their own set of parameters $\,\omega$'s, commonly referred to as \textit{metric fluxes}, which specify the geometry. In addition, topologically non-trivial cycles on the internal space can support constant background fluxes $\,h$'s and $\,f$'s respectively associated with the NS-NS $2$-form gauge potential and the various R-R $p$-form gauge potentials of the higher-dimensional supergravity. In this manner, the lower-dimensional supergravity resulting from the compactification on the group manifold contains more than just the gauge coupling constant $\,g$: it also contains metric and gauge fluxes $(\omega,h,f)$. This opens up the possibility to achieve scale separation between the external AdS spacetime and the internal group manifold upon tuning of the $(\omega,h,f)$-fluxes. However, having parametrically-controlled scale separation is often hindered (if not forbidden) by certain charge conservation conditions, known as tadpole cancellation conditions \cite{Koerber:2007hd}. These take the general form
\begin{equation}
\label{tadpole_cond}
d F_{(8-p)} -  H_{(3)} \wedge F_{(6-p)} = J_{\textrm{O}p} - J_{\textrm{D}p} \ ,
\end{equation}
where $\,H_{(3)}\,$ and $\,F_{(p+1)}\,$ are the NS-NS and R-R gauge fluxes of type II supergravity, and $\,J_{\textrm{O}p}\,$ and $\,J_{\textrm{D}p}\,$ denote the number of charged O$p$-plane and D$p$-brane sources (in the smeared limit) present in the compactification. The number of O$p$-planes in an orientifold compactification is given by the number of fixed points of the associated orientifold involution. This implies an upper bound for the right-hand-side of the tadpole condition (\ref{tadpole_cond}), which can be schematically written as
\begin{equation}
\label{tadpole_cond_schematic}
\omega \, f -  h \, f = J_{\textrm{O}p} - J_{\textrm{D}p} \le J_{\textrm{O}p}  \ .
\end{equation}
This condition makes a parametric control of scale separation more difficult as it often requires a large rescaling of a R-R flux, \textit{i.e.} $f\rightarrow  N \, f$ with $N\rightarrow \infty$,  which clashes with the upper bound in~(\ref{tadpole_cond_schematic}). This problem could a priori be overcome by introducing a compensating rescaling of some metric and/or NS-NS gauge flux, \textit{i.e.} $\,\omega \rightarrow  N^{-1} \, \omega\,$ and/or $\,h \rightarrow  N^{-1} \, h$, but this would now clash with the quantised nature of the fluxes (see \cite{Marchesano:2006ns} for a discussion on the quantisation of metric fluxes). 

Much effort has recently been devoted to investigate scale-separated AdS$_{3}$ flux vacua in $\,\mathcal{N}=1\,$ (\textit{i.e.} $2$ supercharges) type IIA \cite{Farakos:2020phe} (see also \cite{Farakos:2023nms,Farakos:2023wps,Farakos:2025bwf}) and type IIB \cite{Emelin:2021gzx} G$_{2}$ orientifolds. In the $\,\mathcal{N}=1\,$ type IIA compactifications of \cite{Farakos:2020phe}, which include simultaneously O$2$-planes and (intersecting) O$6$-planes and do not allow for metric fluxes, the authors found AdS$_{3}$ flux vacua with small string coupling, large internal volume and scale separation. However, in the $\,\mathcal{N}=1\,$ type IIB compactifications of \cite{Emelin:2021gzx}, which include simultaneously O$9$-planes and (intersecting) O$5$-planes and allow for metric fluxes, the authors observed that AdS$_{3}$ flux vacua could still be found but parametrically-controlled scale separation was obstructed by the quantised nature of the metric fluxes. Building upon \cite{Emelin:2021gzx}, subsequent works \cite{VanHemelryck:2022ynr,Apers:2022vfp} arrived at a similar conclusion: parametrically-controlled scale-separated (classical) AdS$_{3}$ vacua in type IIB seem implausible. Although a systematic scanning of AdS$_{3}$ flux vacua in the $\,\mathcal{N}=1\,$ type IIB compactifications of \cite{Emelin:2021gzx} was actually never performed.

Ref.~\cite{Arboleya:2024vnp}, which we review and extend in these proceedings, marks an initial step towards a systematic scan of flux vacua in the $\,\mathcal{N}=1$, $\,D=3\,$ type IIA and type IIB compactifications of \cite{Farakos:2020phe} and \cite{Emelin:2021gzx}. For a complete charting of flux vacua to be feasible, ref.~\cite{Arboleya:2024vnp} considered a simpler sub-class of type IIA and type IIB orientifold reductions including only \textit{a single type} of O$p$-plane. This implies that the resulting $D=3$ supergravities feature a larger (half-maximal) amount of supersymmetry, namely, they are $\,\mathcal{N}=8\,$ (\textit{i.e.} $16$ supercharges), $\,D=3\,$ supergravities.\footnote{Being framed within the context of half-maximal $\,D=3\,$ supergravity, our program continues the line initiated in \cite{Dibitetto:2011gm} within the context of half-maximal $\,D=4\,$ supergravity, and later extended to half-maximal $\,D=6\,$ \cite{Dibitetto:2019odu} and $\,D \geq 7\,$ \cite{Dibitetto:2012rk,Dibitetto:2015bia} supergravities.} More concretely, ref.~\cite{Arboleya:2024vnp} investigated the following two types of orientifold reductions:

\begin{itemize}

\item Type IIA reductions with O$2$-planes (and D$2$-branes) filling the external $D=3$ spacetime. These are the three-dimensional analogue of the well-studied type IIB reductions with O$3$/D$3$ sources filling an external $D=4$ spacetime \cite{Giddings:2001yu,Kachru:2002he,DeWolfe:2004ns}.

\item Type IIB reductions with O$5$-planes (and D$5$-branes) filling the external $D=3$ spacetime and an internal three-cycle. These are the analogue of the well-studied type IIA reductions with O$6$/D$6$ sources filling an external $D=4$ spacetime and an internal three-cycle \cite{DeWolfe:2005uu,Camara:2005dc,Villadoro:2005cu,Derendinger:2005ph,DallAgata:2009wsi}. 

\end{itemize}

\noindent While type IIA models admit only no-scale Minkowski (Mkw) flux vacua preserving up to $\mathcal{N}=6$ supersymmetry -- consistent with previous findings in \cite{Farakos:2020phe} -- the landscape of type IIB models proves to be very rich. Below we list some of its most salient features.

\begin{enumerate}

\item It contains (non-)supersymmetric Mkw$_{3}$ and AdS$_{3}$ flux vacua (see Table~\ref{Table:flux_vacua_IIB}), all of which remain perturbatively stable within half-maximal supergravity (see Table~\ref{Table:flux_vacua_IIB_masses}).

\item It contains the first examples of (classical) type IIB AdS$_3$ flux vacua that simultaneously feature weak string coupling $\,g_{s}$, large internal volume $\,L_{7} \sim (\textrm{vol}_{7})^{\frac{1}{7}}$, and a parametrically-controlled scale separation between AdS$_{3}$ and the internal seven-dimensional space. Namely,
\begin{equation}
\label{scalings_intro}
g_{s} \sim N^{-a}
\hspace{8mm} , \hspace{8mm}
L_{7} \sim N^{b}
\hspace{8mm} \textrm{ and } \hspace{8mm}
\frac{L}{L_{7}} \sim N^{c}  \ ,
\end{equation}
with $\,a, b, c >0$. $\,N\,$ is a scaling parameter to be taken arbitrarily large (formally $N\rightarrow \infty$) while keeping fluxes quantised and $\,J_{\textrm{O}5} - J_{\textrm{D}5}\,$ fixed in (\ref{tadpole_cond_schematic}). These parametrically-controlled, scale-separated AdS$_{3}$ vacua, labeled \textbf{vac~10} and \textbf{vac~11} in Tables \ref{Table:flux_vacua_IIB} and \ref{Table:flux_vacua_IIB_masses}, are non-supersymmetric, and are re-examined in Section~\ref{sec:scale_sep&int_Deltas_in_vac10,11}. A discussion (not addressed in~\cite{Arboleya:2024vnp}) on the anisotropy of the corresponding seven-dimensional internal spaces is also included. Their supersymmetric cousin found in \cite{VanHemelryck:2025qok} is discussed in Section~\ref{sec:susy_cousin}, where its complete scalar mass spectrum within half-maximal supergravity is presented for the first time and compared with those of the non-supersymmetric \textbf{vac~10} and \textbf{vac~11}.

\item For all the AdS$_{3}$ vacua, the complete spectrum of scalar perturbations is independent of the flux parameters that remain unfixed, and only contains non-negative mass-squared values. Surprisingly (at least to us), most of the AdS$_{3}$ flux vacua -- including the non-supersymmetric, scale-separated \textbf{vac~10} and \textbf{vac~11} -- exhibit integer-valued conformal dimensions $\Delta$'s of the would-be dual CFT$_2$ operators (see Table~\ref{Table:flux_vacua_IIB_masses}), as first noted for DGKT vacua \cite{DeWolfe:2005uu} in \cite{Conlon:2021cjk,Apers:2022tfm}. Furthermore, the specific values of the $\Delta$'s are compatible with polynomial shift symmetries, as noted for DGKT vacua in \cite{Apers:2022vfp}. The absence of negative masses in the spectrum of scalar fluctuations translates, holographically, into the absence of relevant operators in the putative CFT$_2$'s, namely, all $\Delta$'s are $\ge 2$. This provides an arena to (holographically) explore the existence of strongly-coupled dead-end CFT's in two dimensions (see \cite{Nakayama:2015bwa} for a perturbative search for dead-end CFT's in various dimensions).

\end{enumerate}

Although much of the material presented here has appeared in earlier works, especially in ref.~\cite{Arboleya:2024vnp}, the discussion on the internal space anisotropy and the existence of polynomial shift symmetries in the type IIB AdS$_{3}$ flux vacua are, amongst others, original contributions presented in these proceedings.

\section{Type II reductions to half-maximal $D=3$ supergravity}

Following the original construction by Scherk and Schwarz (SS) \cite{Scherk:1979zr}, we will investigate the dimensional reduction of type II supergravity on a (locally) seven-dimensional group manifold associated with a Lie group $\textrm{G}$.

\subsection{Twisted tori}

Group manifolds are sometimes referred to as \textit{twisted tori}, although the term \textit{tori} might be misleading since group manifolds are nothing to do with ordinary tori, \textit{e.g.}, they can have non-zero curvature. Denoting $y^{m}$ $(m=1,\ldots,7)$ the coordinates on the group manifold, let us introduce a (local) basis of one-forms
\begin{equation}
\label{eta_basis}
\eta^{m} = U^{m}{}_{n}(y) \, dy^{n}
\hspace{10mm} \textrm{ with } \hspace{10mm}
U^{m}{}_{n}(y) \in \textrm{G} \ .
\end{equation}
The $U$-twist in (\ref{eta_basis}) renders the one-forms $\eta^{m}$ no longer closed. Instead, they obey the structure equation (or Maurer--Cartan equation)
\begin{equation}
\label{structure_equation}
d\eta^{p} + \tfrac{1}{2} \, \omega_{mn}{}^{p} \, \eta^{m} \wedge \eta^{n} = 0 
\hspace{8mm} \textrm{ with } \hspace{8mm}
\omega_{mn}{}^{p}=(U^{-1})_{m}{}^{r} (U^{-1})_{n}{}^{s} \left( \partial_{r} U^{p}{}_{s} - \partial_{s} U^{p}{}_{r}  \right) \ .
\end{equation}
The isometry generators (Killing vectors) on the group manifold, we denote them by $X_m = (U^{-1})_m{}^n \, \partial_{n}$, satisfy the commutation relations  
\begin{equation}
\label{X_commutators}
[X_m, X_n] = \omega_{mn}{}^{p} \, X_p \ ,
\end{equation}
specified by the structure constants $\,\omega_{mn}{}^{p}\,$ (commonly referred to as \textit{metric fluxes}) of the Lie algebra associated with the Lie group $\textrm{G}$.

Equipped with the basis of one-forms $\,\eta^{m}\,$ in (\ref{eta_basis}), one can then proceed and perform a dimensional reduction of type II supergravity \`a la Kaluza--Klein (KK), \textit{i.e.}, as if the reduction was on a torus\footnote{To emphasise this fact, we will denote the group manifold (or twisted torus) by $\,\mathbb{T}^{7}_{\omega}$.}. The internal part of the Neveu-Schwarz--Neveu-Schwarz (NS-NS) three-form field strength $H_{(3)}$ is expanded as $H_{(3)} = \tfrac{1}{3!} \,H_{mnp} \, \eta^{m} \wedge \eta^{n} \wedge \eta^{p}$ with constant flux parameters $H_{mnp}$. Similarly, the various $p$-form field strengths in the Ramond--Ramond (R-R) sector of type II supergravity are expanded as $F_{(p)} = \tfrac{1}{p!} \,F_{m_{1} \cdots m_{p}} \, \eta^{m_{1}} \wedge \cdots \wedge \eta^{m_{p}}$ with constant flux parameters $F_{m_{1} \cdots m_{p}}$. The outcome of the dimensional reduction is then a three-dimensional supergravity model. As we will see in a moment, the number of preserved supersymmetries in three dimensions depends on the specific choice of metric fluxes $\,\omega_{mn}{}^{p}$ and gauge fluxes $\,H_{mnp}\,$ and $\,F_{m_{1} \cdots m_{p}}$. And a choice of metric and gauge fluxes amounts to a choice of O$p$-plane/D$p$-brane sources in the compactification scheme via the so-called tadpole cancellation conditions.

\subsection{Fluxes, sources and supersymmetry}

Type II string compactifications can also incorporate sources in the compactification scheme: O$p$-planes, D$p$-branes, NS5-branes or KK monopoles amongst other more exotic objects \cite{deBoer:2012ma}. In this work we will consider just O$p$-planes (negative tension) and D$p$-branes (positive tension). These are electrically (magnetically) charged under a $C_{(p+1)}$ ($C_{(7-p)}$) gauge potential of type II supergravity. In both cases, we will consider the sources in the so-called smeared limit\footnote{They are delocalised in the internal directions transverse to their world-volume.} and further ignore their dynamics. Still, being charged objects, they must obey some charge conservation conditions corresponding to Gauss law -- known as tadpole cancellation conditions -- and involving the total charge of the O-planes/D-branes. These tadpole cancellation conditions are of the form (see \textit{e.g.} \cite{Koerber:2007hd})
\begin{equation}
\label{Tadpole_p-form}
d F_{(8-p)} -  H_{(3)} \wedge F_{(6-p)} = J_{\textrm{O}p/\textrm{D}p} \ ,
\end{equation}
where $J_{\textrm{O}p/\textrm{D}p}$ denotes the net current of charged O$p$-plane/D$p$-brane sources in the smeared limit. In addition to the tadpole cancellation conditions (\ref{Tadpole_p-form}), there are two additional conditions
\begin{equation}
\label{Jacobi_&_dH3}
dH_{(3)} = 0
\hspace{15mm} \textrm{ and } \hspace{15mm} 
\omega_{[mn}{}^{r} \, \omega_{p]r}{}^{q} = 0 \ .
\end{equation}
The first condition reflects the absence of NS$5$-branes in our compactification scheme. The second condition is simply the Jacobi identity that follows from (\ref{X_commutators}) and reflects the absence of KK monopoles in our compactification scheme \cite{Villadoro:2007yq}.

The presence of O$p$-plane/D$p$-brane sources breaks supersymmetry: extended objects break some translational symmetries and, therefore, also the supercharges closing into the broken momentum generators in the superalgebra. However, and very importantly, having just \textit{a single type} of O$p$/D$p$ sources (like O$2$/D$2$ in type IIA or O$5$/D$5$ in type IIB) still preserves half of the original $32$ supercharges of type II supergravity. Preserving $32/2=16$ supercharges in three dimensions corresponds to a half-maximal $\mathcal{N}=8$, $D=3$ supergravity. These theories have been systematically constructed in \cite{Marcus:1983hb,Nicolai:2001ac,Deger:2019tem}.

\subsection{$\mathcal{N}=8$, $D=3$ supergravity from type II orientifold reductions}

Half-maximal ($16$ supercharges) supergravities in $D=3$ can be systematically constructed using the so-called embedding tensor formalism \cite{Nicolai:2001ac,Deger:2019tem}. These theories exist in two versions commonly referred to as the \textit{ungauged} and \textit{gauged} cases. 
\\

\noindent Within the context of type II orientifold reductions:
\begin{itemize}

\item The ungauged supergravity arises when reducing type II supergravity on an ordinary (or straight) seven-torus $\mathbb{T}^{7}$ in the absence of background fluxes (both metric and gauge) and in the presence of a single type of O$p$-plane. In this case the conditions in (\ref{Jacobi_&_dH3}) are trivially satisfied whereas the tadpole cancellation condition in (\ref{Tadpole_p-form}) requires
\begin{equation}
\label{J=0_cond}
J_{\textrm{O}p/\textrm{D}p} \equiv J_{\textrm{O}p}-J_{\textrm{D}p} = 0 \ ,
\end{equation}
where $J_{\textrm{O}p}$ is the number of O$p$-planes located at the fixed points of the orientifold involution and $J_{\textrm{D}p}$ is the number of D$p$-branes sitting on top of them in the compactification scheme. Then, the condition (\ref{J=0_cond}) simply forces the net charge of O$p$/D$p$ sources to vanish. Importantly, the half-maximal $D=3$ ungauged supergravity features an $\textrm{SO}(8,8)$ global symmetry that becomes manifest once the vector fields arising from the type II reduction are dualised into scalars.\footnote{This is in the same spirit as the $\textrm{SL}(2)$ Ehlers symmetry that appears in the reduction of standard Einstein gravity to three dimensions \cite{Ehlers:1957zz}.}

\item A gauged supergravity arises when reducing type II supergravity on a group manifold (or twisted seven-torus) with gauge fluxes, and in the presence of a single type of O$p$-planes. In this case the conditions in (\ref{Jacobi_&_dH3}) become non-trivial and the tadpole cancellation condition in (\ref{Tadpole_p-form}) reads
\begin{equation}
\label{J_not_0_cond}
J_{\textrm{O}p/\textrm{D}p} \equiv J_{\textrm{O}p}-J_{\textrm{D}p} \neq 0 \ ,
\end{equation}
for the particular single type of O$p$/D$p$ sources that are allowed in the compactification scheme. The net charge of \textit{any} other type of O$p'$/D$p'$ sources with $p' \neq p$ must vanish by virtue of half-maximal supersymmetry. Once metric and/or gauge fluxes are turned on, the $\textrm{SO}(8,8)$ global symmetry of the ungauged $D=3$ supergravity gets broken.

\end{itemize}

The field content of the resulting $\mathcal{N}=8$, $D=3$ supergravity is fixed (and the same for both the ungauged and gauged cases). There is the supergravity multiplet (which includes the spacetime metric and eight gravitini) which is coupled to $n=8$ matter multiplets (each of which consists of eight scalar fields and eight spin-$\frac{1}{2}$ fields). The bosonic part of the Lagrangian is therefore of Einstein-scalar type and takes the form (in the conventions of \cite{Deger:2019tem})
\begin{equation}
\label{L_bosonic}
e^{-1} \, \mathcal{L}_{\textrm{bos}} = - \frac{R}{4} - \frac{1}{32} \, \textrm{Tr}\left(\partial_\mu M \, \partial^{\mu} M^{-1}\right) - V \ ,
\end{equation}
where $\,e=\sqrt{-|g|}\,$ and $\,M \in \textrm{SO}(8,8)\,$ is a matrix that depends on the $\,8 \times 8=64\,$ scalar fields of the half-maximal supergravity. The scalars in the theory serve as coordinates in the coset space
\begin{equation}
\label{scalar_geometry_N=8}
{\mathcal{M}}_{\textrm{scal}}^{\mathcal{N}=8} = \frac{{\rm SO}(8,8)}{{\rm SO}(8) \times {\rm SO}(8)}  \ ,
\end{equation}
so the scalar-dependent matrix $\,M\,$ can be parameterised as
\begin{equation}
\label{M_parameterisation_N=8}
M = \left( 
\begin{matrix}  \boldsymbol{g} & -\boldsymbol{g} \, \boldsymbol{b} \\ 
\boldsymbol{b} \, \boldsymbol{g} & \boldsymbol{g}^{-1} - \boldsymbol{b} \, \boldsymbol{g} \, \boldsymbol{b}
\end{matrix} 
\right) \in \textrm{SO}(8,8)
\hspace{8mm} \textrm{ with } \hspace{8mm}
\boldsymbol{g} = \boldsymbol{e} \, \boldsymbol{e}^{T} \ ,
\end{equation}
in terms of $\,8 \times 8\,$ matrices $\,\boldsymbol{e} \in \textrm{GL}(8)/\textrm{SO}(8)\,$ and $\,\boldsymbol{b}=-\boldsymbol{b}^{T}\,$ containing $\,36\,$ and $\,28\,$ scalars, respectively. These scalars originate from the purely internal components of the ten-dimensional fields ($g_{mn}$, $B_{mn}$, etc.), as well as from components with one leg along the external spacetime ($g_{m\mu}$, $B_{m\mu}$, etc.) upon dualisation into scalars in $D=3$. Finally, when fluxes are turned on in the compactification scheme (\textit{i.e.} in the gauged case), they induce the scalar potential in (\ref{L_bosonic}) which takes the schematic form
\begin{equation}
\label{V_N=8}
V(\Theta \, ; M) = g^{2} \,  \Theta  \, \Theta \left( M^{4} + M^{3} + \dots \right) \ ,
\end{equation}
where $\,g\,$ is the gauge coupling constant and $\,\Theta\,$ is an object known as the \textit{embedding tensor} \cite{Nicolai:2001ac}.\footnote{We will set the gauge coupling constant $\,g\,$ of the three-dimensional half-maximal gauged supergravity to $\,g=1$. This amounts to a global rescaling $\,\Theta \rightarrow \Theta/g\,$ of the $\Theta$-tensor in (\ref{V_N=8}).} Importantly, $\,\Theta\,$ encodes, in the $D=3$ supergravity, the various fluxes (both metric and gauge) in the higher-dimensional compactification scheme. We note in passing that the scalar potential in (\ref{V_N=8}) is quadratic in the embedding tensor $\,\Theta\,$ (equivalently, in the flux parameters) whereas it involves higher powers of the scalars through the scalar-dependent matrix $\,M\,$.

\subsection{The fluxes $\leftrightarrow$ $\Theta$ correspondence}

The embedding tensor $\,\Theta\,$ in the half-maximal $D=3$ gauged supergravity encodes the various fluxes, both metric and gauge, in the type II compactification scheme. This tensor has a decomposition into irreducible representations (irrep's) of $\,\textrm{SO}(8,8)\,$ of the form
\begin{equation}
\label{ET_N=8}
\Theta_{MN|PQ} = \theta_{MNPQ} + 2 \left( \eta_{M[P} \theta_{Q]N} - \eta_{N[P} \theta_{Q]M} \right) + 2 \, \eta_{M[P} \eta_{Q]N} \theta \ ,
\end{equation}
where
\begin{equation}
\label{ET_irreps}
\theta_{MNPQ}=\theta_{[MNPQ]} \in \mathbf{1820}
\hspace{6mm} , \hspace{6mm}
\theta_{MN}=\theta_{(MN)} \in \mathbf{135} \,\,\,\,(\textrm{traceless})
\hspace{6mm} \textrm{ and } \hspace{6mm}
\theta \in \mathbf{1} \ .
\end{equation}
The $\,\eta_{MN}\,$ matrix entering (\ref{ET_N=8}) is the non-degenerate $\textrm{SO}(8,8)$ invariant matrix that is used to raise/lower vector indices $\,M,N=1,\ldots,16$. Finally, the consistency of the gauged supergravity requires a so-called \textit{quadratic constraint} on the embedding tensor that can be written as \cite{Eloy:2024lwn}
\begin{equation}
\label{QC_N=8}
\Theta_{KL|[M}{}^{R} \, \Theta_{N]R|PQ} + \Theta_{KL|[P}{}^{R} \, \Theta_{Q]R|MN} = 0 \ .
\end{equation}

The precise dictionary between flux parameters and components of $\,\Theta\,$ depends on the type IIA/IIB starting point as well as on the specific orientifold action $\,\mathcal{O}_{\mathbb{Z}_{2}}\,$ associated with the O$p$-plane. The orientifold action $\,\mathcal{O}_{\mathbb{Z}_{2}}\,$ combines worldsheet and target space transformations, including a worldsheet parity transformation $\,\Omega_{P}$, a left-moving fermion number projector $\,(-1)^{F_{L}}\,$ (sometimes unnecessary \cite{Koerber:2007hd}), and an internal space involution $\,\sigma_{\textrm{O}p}$.\footnote{If $\,\sigma_{\textrm{O}p}\,$ leaves some sub-manifold of the internal space invariant, its product with the external spacetime is identified with the O$p$-plane. The integer $\,p\,$ denotes the total number of spatial dimensions threaded by the O$p$-plane.} The orientifold action $\mathcal{O}_{\mathbb{Z}_{2}}$ on the string theory side is precisely the $\,\mathbb{Z}_{2}\,$ discrete symmetry that group-theoretically truncates from maximal to half-maximal supergravity in $\,D=3$.

\subsection{An example: type IIA with O$2$/D$2$ sources}

Let us consider the case of type IIA reductions with orientifold O$2$-planes (and possibly D$2$-branes). Three-dimensional Lorentz invariance requires these sources to be located at
\begin{equation}
\label{O2_location}
\begin{array}{lll|lc|lc|lc|c}
x^{0} & x^{1} & x^{2} & \eta^{1}  & \eta^{2} & \eta^{3}  & \eta^{4}& \eta^{5} & \eta^{6} & \eta^{7}   \\
\hline
\times & \times &\times & &  &  &  &  &  &  
\end{array}   
\end{equation}
thus filling the external spacetime completely. The internal target space involution $\,\sigma_{\textrm{O}2}\,$ reflects all the internal basis elements transverse to the O$2$-plane, \textit{ i.e.}, $\,\sigma_{\textrm{O}2}: \, \eta^{m} \rightarrow -\eta^{m}$, with $\,m=1,\ldots,7$. Therefore, $\,\sigma_{\textrm{O}2}\,$ is compatible with an $\,\textrm{SL}(7) \subset \textrm{SO}(8,8)\,$ covariant formulation of the $D=3$ supergravity.

\begin{table}[t]
\begin{center}
\renewcommand{\arraystretch}{1.7}
\begin{tabular}{|c|c|c|c|c|c|c|c|c|}
\hline
Fields &    $e_{n}{}^{p}$ &  $C_{(1)}$ &  $\Phi$ & $C_{(3)}$ &   $B_{(2)}$ & $B_{(6)}$ & $C_{(5)}$ \\
\hline
$\Omega_{P}$ & $+$  & $-$  & $+$ & $+$ & $-$ & $+$ &  $-$  \\
\hline
$\sigma_{\textrm{O}2}$  & $+$  & $-$ & $+$ & $-$ & $+$ & $+$ & $-$    \\
\hline
$\mathcal{O}_{\mathbb{Z}_{2}}$  & $+$  & $+$ & $+$ & $-$ &  $-$ & $+$ & $+$  \\
\hline\hline
Fluxes &    $\omega_{mn}{}^{p}$ &  $F_{(2)}$ &  $H_{(1)}$ & $F_{(4)}$ &   $H_{(3)}$ & $H_{(7)}$ & $F_{(6)}$ \\
\hline  
$\mathcal{O}_{\mathbb{Z}_{2}}$  & $-$  & $-$ & $-$ & $+$ & $+$ & $-$ & $-$  \\
\hline
\end{tabular}
\caption{Even ($+$) \textit{vs.} odd ($-$) nature of type~IIA fields and fluxes under the O$2$-plane orientifold action $\mathcal{O}_{\mathbb{Z}_{2}} = \Omega_{P} \, \sigma_{\textrm{O}2}$.} 
\label{Table:O2_fields_fluxes}
\end{center}
\end{table}

The full orientifold action $\mathcal{O}_{\mathbb{Z}_{2}} = \Omega_{P} \, \sigma_{\textrm{O}2}$ acts on the type~IIA fields and fluxes as shown in Table~\ref{Table:O2_fields_fluxes}. From there,\footnote{The Romans mass parameter $F_{(0)}$ is not included in Table~\ref{Table:O2_fields_fluxes} as it does not originate from a type IIA gauge potential.} the $\mathcal{O}_{\mathbb{Z}_{2}}$-even fields are identified with
\begin{equation}
\label{Fields&Fluxes_bosonic_IIA}
\begin{array}{lcl}
\textrm{Scalars} & : & e_{m}{}^{n} \in \frac{\textrm{GL}(7)}{\textrm{SO}(7)}
\hspace{3mm} , \hspace{3mm}
C_{(1)}
\hspace{3mm} , \hspace{3mm}
\Phi
\hspace{3mm} , \hspace{3mm}
B_{(6)}
\hspace{3mm} , \hspace{3mm}
C_{(5)} \ , \\[2mm]
\textrm{Fluxes} & : & 
H_{(3)}
\hspace{3mm} , \hspace{3mm}
F_{(4)} 
\hspace{3mm} , \hspace{3mm}
F_{(0)} \ ,
\end{array}
\end{equation}
where $\,e_{m}{}^{n}\,$ is the internal sieben-bein. Notice that the $\,64\,$ scalars of half-maximal $\,D=3\,$ supergravity are recovered. The various gauge fluxes in (\ref{Fields&Fluxes_bosonic_IIA}) are placed in the embedding tensor irreps of (\ref{ET_irreps}) as
\begin{equation}
\label{ET_flux_dictionary_IIA_O2}
\theta^{mnpq} = \tfrac{1}{3!} \, \varepsilon^{mnpqrst} H_{rst}
\hspace{5mm} , \hspace{5mm}
\theta^{mnp8} = \tfrac{1}{4!} \, \varepsilon^{mnpqrst} F_{qrst}
\hspace{5mm} , \hspace{5mm}
\theta^{88} = F_{(0)} \ .
\end{equation}
Since the metric fluxes $\,\omega_{mn}{}^{p}\,$ are projected out by the orientifold action, it follows that $\,d\eta^{m}=0\,$ in (\ref{structure_equation}) and the conditions in (\ref{Jacobi_&_dH3}) are automatically satisfied for constant flux parameters $\,H_{mnp}$. However, there is a non-trivial flux-induced tadpole for O$6$/D$6$ sources which must vanish since such sources are forbidden in our half-maximal setup (recall that only O$2$/D$2$ sources are allowed). Setting $\,p=6\,$ in (\ref{Tadpole_p-form}), and noticing that $\,F_{(2)}\,$ is projected out by the orientifold action (see Table~\ref{Table:O2_fields_fluxes}), one arrives at the tadpole cancellation condition
\begin{equation}
\label{Tadpole_IIA_O2}
H_{(3)} \, F_{(0)} = 0 \ .
\end{equation}
This tadpole cancellation condition, which is quadratic in the flux parameters, precisely matches the quadratic constraint that follows from plugging (\ref{ET_flux_dictionary_IIA_O2}) into (\ref{ET_N=8}) and (\ref{QC_N=8}).

Equipped with the precise dictionary between gauge fluxes $\left\lbrace H_{mnp} \,,\, F_{mnpq} \,,\, F_{(0)} \right\rbrace$ and $\Theta$-tensor components in (\ref{ET_flux_dictionary_IIA_O2}), one can proceed to compute the scalar potential (\ref{V_N=8}) and analyse its vacuum structure, namely, its set of critical points corresponding to maximally symmetric solutions of the $D=3$ supergravity. However, the tadpole condition (\ref{Tadpole_IIA_O2}) prevents $F_{(0)}$ and $H_{(3)}$ from being simultaneously turned on if the compactification scheme includes
O$2$-planes and is to preserve half-maximal supersymmetry. One is then left with two possible situations:
\begin{itemize}

\item[$i)$] $F_{(0)}=0$ : In this case the compactification includes the gauge fluxes $\,\left\lbrace H_{(3)} \,,\, F_{(4)} \right\rbrace$. The scalar potential becomes a sum of squares and possesses an overall modulus corresponding to a (unstabilised) massless ``no-scale" direction \cite{Farakos:2020phe}. The critical locus of $\,V$, we denote it $\,M_{0}$, only contains Minkowski vacua (\textit{i.e.} $V_{0}=0$). The full spectrum of scalar fluctuations around $\,M_{0}\,$ and the number of preserved supersymmetries were computed in \cite{Arboleya:2024vnp} as a function of the flux parameters $\,\left\lbrace H_{(3)} \,,\, F_{(4)} \right\rbrace$. Remarkably, the Minkowski vacua can preserve from $\,\mathcal{N}=0\,$ up to $\,\mathcal{N}=6\,$ supersymmetries depending on the choice of flux parameters.

\item[$ii)$] $H_{(3)}=0$ : In this case the compactification includes the gauge fluxes $\,\left\lbrace F_{(0)} \,,\, F_{(4)} \right\rbrace$. The extremisation of $\,V\,$ then requires that either all the fluxes are zero, thus trivialising $V$, or the moduli fields parameterising the size of the internal cycles go to infinity, thus rendering the compactification singular. Therefore, there are no vacua in this case.

\end{itemize}

\section{Type IIB AdS$_{3}$ flux vacua with O$5$/D$5$ sources}

Let us now consider type IIB reductions with O$5$-planes (and possibly D$5$-branes) filling the external spacetime and an internal three-cycle which we choose it to be $\,\eta^{246} \equiv \eta^{2} \wedge \eta^{4} \wedge \eta^{6}\,$ (without loss of generality). The O$5$/D$5$ sources are therefore located as
\begin{equation}
\label{O5_location}
\begin{array}{lll|lc|lc|lc|c}
x^{0} & x^{1} & x^{2} & \eta^{1}  & \eta^{2} & \eta^{3}  & \eta^{4}& \eta^{5} & \eta^{6} & \eta^{7}   \\
\hline
\times & \times &\times & & \times &  & \times &  & \times &  
\end{array}   
\end{equation}
Out of the basis elements $\,\eta^{m}\,$ in (\ref{eta_basis}), the internal target space involution $\,\sigma_{\textrm{O}5}\,$ reflects the basis elements $\,\eta^{\hat{a}}\,$ $\,(\,\hat{a}=1,3,5,7\,)$, \textit{ i.e.}, $\,\sigma_{\textrm{O}5}: \, \eta^{\hat{a}} \rightarrow -\eta^{\hat{a}}$, and leaves invariant the basis elements $\,\eta^{i}\,$ $\,(\,i=2,4,6\,)$, \textit{ i.e.}, $\,\sigma_{\textrm{O}5}: \, \eta^{i} \rightarrow \eta^{i}$. As a result, $\,\sigma_{\textrm{O}5}\,$ is only compatible with an $\,\textrm{SL}(4) \times \textrm{SL}(3) \subset \textrm{SO}(8,8)\,$ covariant formulation of the $D=3$ supergravity.

\begin{table}[t]
\begin{center}
\renewcommand{\arraystretch}{1.7}
\begin{tabular}{|c|c|c|c|}
\hline
Fields & $\Omega_{P}$   &  $\sigma_{\textrm{O}5}$ & $\mathcal{O}_{\mathbb{Z}_{2}}$  \\
\hline
\hline
$e_{\hat{a}}{}^{\hat{b}} \in \frac{\textrm{GL}(4)}{\textrm{SO}(4)} \, , \, e_{i}{}^{j} \in \frac{\textrm{GL}(3)}{\textrm{SO}(3)}$ & \multirow{2}{*}{$+$}  & $+$  & $+$  \\
\cline{1-1}\cline{3-4}
$e_{\hat{a}}{}^{i}  \, , \, e_{i}{}^{\hat{a}}$ &  & $-$  & $-$ \\
\hline
\hline
$\Phi$ & $+$ & $+$  & $+$ \\
\hline
\hline
$B_{ij}\, , \, B_{\hat{a}\hat{b}}$ & \multirow{2}{*}{$-$}  & $+$  & $-$  \\
\cline{1-1}\cline{3-4}
$B_{i\hat{a}}$ &  & $-$  & $+$ \\
\hline
\hline
$B_{ij\hat{a}\hat{b}\hat{c}\hat{d}}$ & \multirow{2}{*}{$-$}  & $+$  & $-$  \\
\cline{1-1}\cline{3-4}
$B_{ijk\hat{a}\hat{b}\hat{c}}$ &  & $-$  & $+$ \\
\hline
\end{tabular}
\hspace{10mm}
\begin{tabular}{|c|c|c|c|}
\hline
Fields & $\Omega_{P}$   &  $\sigma_{\textrm{O}5}$ & $\mathcal{O}_{\mathbb{Z}_{2}}$  \\
\hline
\hline
$C_{\hat{a}\hat{b}\hat{c}\hat{d}} \, , \, C_{ij\hat{a}\hat{b}}$ & \multirow{2}{*}{$-$}  & $+$  & $-$  \\
\cline{1-1}\cline{3-4}
$C_{i\hat{a}\hat{b}\hat{c}} \, , \, C_{ijk\hat{a} }$ &  & $-$  & $+$ \\
\hline
\hline
$C_{(0)}$ & $-$ & $+$  & $-$ \\
\hline
\hline
$C_{ij}\, , \, C_{\hat{a}\hat{b}}$ & \multirow{2}{*}{$+$}  & $+$  & $+$  \\
\cline{1-1}\cline{3-4}
$C_{i\hat{a}}$ &  & $-$  & $-$ \\
\hline
\hline
$C_{ij\hat{a}\hat{b}\hat{c}\hat{d}}$ & \multirow{2}{*}{$+$}  & $+$  & $+$  \\
\cline{1-1}\cline{3-4}
$C_{ijk\hat{a}\hat{b}\hat{c}}$ &  & $-$  & $-$ \\
\hline
\end{tabular}
\caption{Even ($+$) \textit{vs.} odd ($-$) nature of type~IIB fields under the O$5$-plane orientifold action $\mathcal{O}_{\mathbb{Z}_{2}} = \Omega_{P} \, \sigma_{\textrm{O}5}$. The $\mathcal{O}_{\mathbb{Z}_{2}}$-even fields add up to $\,64-3=61\,$ scalars. The missing three scalars in half-maximal $\,D=3\,$ supergravity are dual to the vectors $\,e_{\mu}{}^{i}$.} 
\label{Table:O5_fields}
\end{center}
\end{table}

\begin{table}[t]
\begin{center}
\renewcommand{\arraystretch}{1.8}
\begin{tabular}{|c|c|}
\hline
Fluxes &  $\Theta$-irrep $\leftrightarrow$ Flux component \\
\hline
\hline
\multirow{3}{*}{$\omega$} & $\theta^{ij8}{}_{k}=\omega_{ij}{}^{k}$  \\
\cline{2-2}
 & $\theta^{\hat{a}\hat{b}ij} = \frac{1}{2!} \, \epsilon^{\hat{a}\hat{b}\hat{c}\hat{d}} \, \epsilon_{ijk}\, \omega_{\hat{c}\hat{d}}{}^{k}  $  \\
\cline{2-2}
 & $\theta^{i\hat{b}8}{}_{\hat{a}} = \omega_{\hat{a}i}{}^{\hat{b}}$ \\
\hline
\hline
\multirow{2}{*}{$H_{(3)}$} & $ \theta^{ijk\hat{d}} = \frac{1}{3!} \, \epsilon_{ijk} \, \epsilon^{\hat{a}\hat{b}\hat{c}\hat{d}} \, H_{\hat{a}\hat{b}\hat{c}}$ \\
\cline{2-2}
 & $\theta^{ij8}{}_{\hat{c}} = H_{ij\hat{c}}$ \\
\hline
\end{tabular}
\hspace{10mm}
\begin{tabular}{|c|c|}
\hline
Fluxes &   $\Theta$-irrep  $\leftrightarrow$ Flux component  \\
\hline
\hline
$F_{(5)}$ & $\theta^{\hat{d}ij8} = \frac{1}{3!} \, \epsilon^{\hat{a}\hat{b}\hat{c}\hat{d}} \, F_{\hat{a}\hat{b}\hat{c}ij}$ \\
\hline
\hline
\multirow{2}{*}{$F_{(3)}$} & $\theta^{88}  = - \frac{1}{3!} \, \epsilon^{ijk} \, F_{ijk}  $  \\
\cline{2-2}
 & $\theta^{\hat{a}\hat{b}k8} = \frac{1}{2!} \, \epsilon^{\hat{a}\hat{b}\hat{c}\hat{d}} \, F_{\hat{c}\hat{d}k} $  \\
\hline
\hline
$F_{(7)}$ & $ \theta^{ijk8} =  \frac{1}{4!} \, \epsilon^{\hat{a}\hat{b}\hat{c}\hat{d}} \, F_{\hat{a}\hat{b}\hat{c}\hat{d}ijk}$ \\
\hline
\end{tabular}
\caption{Type~IIB (even) fluxes allowed by the orientifold action $\,\mathcal{O}_{\mathbb{Z}_{2}} = \Omega_{P} \, \sigma_{\textrm{O}5}\,$ and their identification with embedding tensor irreps in (\ref{ET_irreps}).} 
\label{Table:O5_fluxes}
\end{center}
\end{table}

The full orientifold action $\,\mathcal{O}_{\mathbb{Z}_{2}} = \Omega_{P} \, \sigma_{\textrm{O}5}\,$ acts on the type~IIB fields as displayed in Table~\ref{Table:O5_fields}. The associated field strengths are obtained upon acting with ($U^{-1}$-twisted) internal derivatives and the result is displayed in Table~\ref{Table:O5_fluxes}. Since some components of the metric flux $\,\omega_{mn}{}^{p}\,$ are now allowed by the orientifold action, it turns out that $\,d\eta^{m} \neq 0\,$ in (\ref{structure_equation}) and the two conditions in (\ref{Jacobi_&_dH3}) become non-trivial. Recall that they correspond to the absence of NS$5$-branes and KK monopoles in the compactification scheme. In addition, there are non-trivial flux-induced tadpoles for O$3$/D$3$ sources as well as O$5$/D$5$ sources \textit{different from those in (\ref{O5_location})}. We must ensure that they vanish, since having additional types of sources is forbidden by the half-maximal supersymmetry of the compactification scheme. Setting $\,p=3\,$ and $\,p=5\,$ in (\ref{Tadpole_p-form}) respectively yields non-trivial tadpole cancellation conditions of the form
\begin{equation}
\label{Tadpole_IIB_O5}
d F_{(5)} -  H_{(3)} \wedge F_{(3)} = 0
\hspace{10mm} \textrm{ and } \hspace{10mm}
d F_{(3)} \Big|_{\Sigma_{(4)}} = 0 \ ,
\end{equation}
where $\,\Sigma_{(4)}\,$ denotes \textit{any} four-form basis element different from $\,\eta^{1357}$. We have verified that the entire set of conditions we must impose on the fluxes precisely matches the set of quadratic constraints that follows from (\ref{QC_N=8}). In particular,
\begin{equation}
\label{Tadpole_O5/D5_allowed}
d F_{(3)} \Big|_{\eta^{1357}} = \textrm{unrestricted} \ ,
\end{equation}
in agreement with the fact that O$5$/D$5$ sources of the type (\ref{O5_location}) are compatible with half-maximal supersymmetry. This represents a non-trivial check of the flux/$\Theta$-dictionary presented in Table~\ref{Table:O5_fluxes}.

Using the precise dictionary between fluxes and $\Theta$-tensor components from Table~\ref{Table:O5_fluxes}, one can compute the scalar potential (\ref{V_N=8}) for type IIB reductions with O$5$/D$5$ sources. The presence of metric fluxes introduces additional complexity, making the scalar potential more intricate than its type IIA counterpart with O$2$/D$2$ sources. Furthermore, when exploring the vacuum landscape, the extremisation equations must be supplemented with the conditions in (\ref{Jacobi_&_dH3}) and (\ref{Tadpole_IIB_O5}). To tackle this problem, we will employ the algebraic geometry software \textsc{Singular} \cite{DGPS}.

\subsection{RSTU-model}

The number of flux parameters in Table~\ref{Table:O5_fluxes} permitted by the orientifold action $\,\mathcal{O}_{\mathbb{Z}_{2}} = \Omega_{P} \, \sigma_{\textrm{O}5}\,$ is too large to allow for an exhaustive classification of flux vacua. Moreover, the presence of $\,64\,$ scalars in half-maximal $\,D=3\,$ supergravity makes the extremisation problem highly complex, if not intractable. To have a more tractable system, we will restrict to the subset of fluxes and fields that are invariant under a specific $\,\textrm{SO}(3)\,$ symmetry embedded into the duality group as
\begin{equation}
\label{SO(3)_embedding}
\textrm{SO}(3)
\subset 
\textrm{SO}(3) \times \textrm{SO}(2,2) \times  \textrm{SO}(2,2)
\subset
 \textrm{SO}(6,6) \times \textrm{SO}(2,2) 
\subset
\textrm{SO}(8,8) \ .
\end{equation}
From the commutant of $\,\textrm{SO}(3)\,$ inside $\,\textrm{SO}(8,8)$, we identify the scalar geometry associated with the $\,\textrm{SO}(3)\,$ invariant sector of half-maximal $D=3$ supergravity with the coset space\footnote{We have used that $\,\textrm{SO}(2,2) \sim \textrm{SL}(2) \times \textrm{SL}(2)$.}
\begin{equation}
\label{scalar_geometry_RSTU}
{\mathcal{M}}_{\textrm{scal}} \,\, = \,\, \left[ \frac{\textrm{SL}(2)}{\textrm{SO}(2)}\right]^{4} \,\, \subset \,\, {\mathcal{M}}_{\textrm{scal}}^{\mathcal{N}=8}  \,  \ .
\end{equation}
Each $\,\textrm{SL}(2)/\textrm{SO}(2)\,$ factor in (\ref{scalar_geometry_RSTU}) describes a copy of the Poincar\'e half-plane and is parameterised by a complex scalar. We will denote the four complex scalars $R$, $S$, $T$ and $U$ and will refer to this $\,\mathcal{N}=2\,$ subsector of half-maximal $\,D=3\,$ supergravity as the RSTU-model.\footnote{Using representation theory, one can verify that $\,2\,$ out of the $\,8\,$ gravitini of half-maximal supergravity are invariant under the $\,\textrm{SO}(3)\,$ in (\ref{SO(3)_embedding}). Therefore, the RSTU-model describes an $\,\mathcal{N}=2\,$ subsector of half-maximal $\,D=3\,$ supergravity. This is the three-dimensional analogue of the extensively studied $\,\mathcal{N}=1\,$ STU-model in four dimensions \cite{Kachru:2002he,Derendinger:2004jn,DeWolfe:2004ns,Camara:2005dc,Villadoro:2005cu,Derendinger:2005ph,DeWolfe:2005uu,Aldazabal:2006up,Dibitetto:2011gm}.} The four complex scalars enter the $\,8 \times 8\,$ $\,\boldsymbol{e}\,$ and $\,\boldsymbol{b}\,$ matrices in (\ref{M_parameterisation_N=8}) as
\begin{equation}
\label{coset_blocks_Z2^3}
\boldsymbol{e} \,\, = \,\, \left( 
\begin{matrix}  
\boldsymbol{e}_{TU} \otimes \mathbb{I}_{3}  & 0 \\ 
0 & \boldsymbol{e}_{RS}
\end{matrix} 
\right)
\hspace{8mm} , \hspace{8mm}
\boldsymbol{b} \,\, = \,\,
\left( 
\begin{matrix}  
\boldsymbol{b}_{T} \otimes \mathbb{I}_{3} & 0 \\ 
0 & \boldsymbol{b}_{R}
\end{matrix}
\right)  \ ,
\end{equation}
with
\begin{equation}
\label{coset_blocks_2x2_1}
\boldsymbol{e}_{TU} \,\, = \,\, \frac{1}{\sqrt{\textrm{Im}T \, \textrm{Im}U}} \left( 
\begin{matrix}  
1 & 0 \\ 
\textrm{Re}U & \textrm{Im}U  
\end{matrix} 
\right)
\hspace{4mm} , \hspace{4mm}
\boldsymbol{e}_{RS} \,\, = \,\, \frac{1}{\sqrt{\textrm{Im}R \, \textrm{Im}S}} \left( 
\begin{matrix}  
1 & 0 \\ 
\textrm{Re}S & \textrm{Im}S  
\end{matrix} 
\right) \ ,
\end{equation}
and
\begin{equation}
\label{coset_blocks_2x2_2}
\,\,\,\, \boldsymbol{b}_{T} \,\, = \,\,
\left( 
\begin{matrix}  
0 & \textrm{Re}T  \\ 
-\textrm{Re}T & 0 
\end{matrix}
\right)  
\hspace{5mm} , \hspace{5mm}
\boldsymbol{b}_{R} \,\, = \,\,
\left( 
\begin{matrix}  
0 & \textrm{Re}R \\ 
-\textrm{Re}R & 0 
\end{matrix}
\right)  \ .
\end{equation}
Note that $\,\textrm{Im}R \, , \, \textrm{Im}S  \, , \,  \textrm{Im}T  \, , \,  \textrm{Im}U > 0\,$ for the complex scalars to parameterise four copies of the Poincar\'e (upper) half-plane.

In order to describe the set of $\,\textrm{SO}(3)\,$ invariant flux parameters, we need to split the $\,\textrm{SL}(4)\,$ index as $\,\hat{a}=(a,7)\,$ with $\,a=1,3,5\,$. Then, the set of $\,\textrm{SO}(3)\,$ invariant fluxes consists of metric fluxes of the form
\begin{equation}
\label{metric_flux_SO(3)}
\begin{array}{lclclc}
\omega_{ab}{}^{k}=\omega_{1} \, \epsilon_{ab}{}^{k}
& \hspace{5mm} , & \hspace{5mm} 
\omega_{7a}{}^{i}=\omega_{2} \, \delta_{a}^{i}
& \hspace{5mm} , & \hspace{5mm} 
\omega_{i7}{}^{a}=\omega_{3} \, \delta_{i}^{a} & , \\[2mm]
\omega_{ai}{}^{7}=\omega_{4} \, \delta_{ai}
& \hspace{5mm} , & \hspace{5mm} 
\omega_{aj}{}^{c}=-\omega_{5} \, \epsilon_{aj}{}^{c}
& \hspace{5mm} , & \hspace{5mm} 
\omega_{ij}{}^{k}=\omega_{6} \, \epsilon_{ij}{}^{k} & ,
\end{array}
\end{equation}
as well as NS-NS fluxes
\begin{equation}
\label{gauge_flux_SO(3)_H}
\begin{array}{c}
H_{abc}= h_{31}  \, \epsilon_{abc}
\hspace{10mm} , \hspace{10mm} 
H_{aij}= h_{32}  \, \epsilon_{aij}  \ ,
\end{array}
\end{equation}
and R-R fluxes
\begin{equation}
\label{gauge_flux_SO(3)_F}
\begin{array}{c}
F_{ijk}= -f_{31}  \, \epsilon_{ijk}
\hspace{8mm} , \hspace{8mm} 
F_{ia7}= f_{32}  \, \delta_{ia}
\hspace{8mm} , \hspace{8mm} 
F_{ibc}= f_{33}  \, \epsilon_{ibc} \ , \\[4mm]
F_{abij7}= -f_{5}  \, \delta_{ai} \, \delta_{bj}
\hspace{8mm} , \hspace{8mm} 
F_{abcijk7}= f_{7}  \, \epsilon_{abc} \, \epsilon_{ijk} \ ,
\end{array}
\end{equation}
where a proper antisymmetrisation of indices is understood in \eqref{metric_flux_SO(3)}-\eqref{gauge_flux_SO(3)_F}. The RSTU-model thus involves $6+2+5=13$ arbitrary flux parameters and four complex scalars. Our goal will be to chart its landscape of vacua.

\subsubsection{Charting the landscape of the RSTU-model}

The scalar potential of the RSTU-model comes along with two symmetries that involve the simultaneous transformation of moduli fields and flux parameters:
\begin{itemize}

\item[$i)$] The first symmetry is a rescaling of (each of) the imaginary parts of the complex scalars accompanied by an appropriate rescaling of the flux parameters.

\item[$ii)$] The second symmetry is a shift of the axions $\,\textrm{Re}T\,$ and $\,\textrm{Re}R\,$ in (\ref{coset_blocks_2x2_2}) accompanied by an appropriate shift of the flux parameters.

\end{itemize}

The combination of these two symmetries of the potential allows us to search for vacuum expectation values (VEV's) of the form
\begin{equation}
\label{GTTO_1}
\left\langle  R \, \right\rangle = i
\hspace{8mm} , \hspace{8mm}
\left\langle  S \right\rangle = \left\langle  \textrm{Re}S \right\rangle + i
\hspace{8mm} , \hspace{8mm}
\left\langle  T \right\rangle = i
\hspace{8mm} , \hspace{8mm}
\left\langle  U \right\rangle = \left\langle  \textrm{Re}U \right\rangle + i  \ ,
\end{equation}
while maintaining generality. Nonetheless, to establish a connection with the $\,\mathcal{N}=1\,$ type IIB orientifold models of \cite{Emelin:2021gzx}, we will further restrict to vacua without axions, namely,
\begin{equation}
\label{GTTO_2}
\left\langle  \textrm{Re}S \right\rangle = \left\langle  \textrm{Re}U\ \right\rangle = 0 \ .
\end{equation}
This corresponds to searching for vacua at the origin of the moduli space, \textit{i.e.}, $M_{0}=\mathbb{I}$, which can afterwards be moved to other positions in field space using the scaling and shift symmetries discussed above. The complete set of extrema of the scalar potential at the origin of moduli space can be obtained employing the algebraic geometry software \textsc{Singular} \cite{DGPS}. The result is summarised in Table~\ref{Table:flux_vacua_IIB}.

\begin{table}[t!]
\begin{center}
\scalebox{0.85}{
\renewcommand{\arraystretch}{1.5}
\begin{tabular}{!{\vrule width 1.5pt}l!{\vrule width 1pt}c!{\vrule width 1pt}c!{\vrule width 1pt}cccccc!{\vrule width 1pt}cc!{\vrule width 1pt}ccc!{\vrule width 1pt}c!{\vrule width 1pt}c!{\vrule width 1.5pt}}
\Xcline{4-16}{1.5pt}
\multicolumn{3}{c!{\vrule width 1pt}}{}& \multicolumn{6}{c!{\vrule width 1pt}}{$\omega$} & \multicolumn{2}{c!{\vrule width 1pt}}{$H_{(3)}$} & \multicolumn{3}{c!{\vrule width 1pt}}{$F_{(3)}$} & \multicolumn{1}{c!{\vrule width 1pt}}{$F_{(5)}$} & \multicolumn{1}{c!{\vrule width 1pt}}{$F_{(7)}$}\\ 
\noalign{\hrule height 1.5pt}
     \hspace{2mm}  ID & Type &  SUSY & $\omega_{1}$ & $ \omega_{2}$ & $  \omega_{3}$ & $\omega_{4}$ & $ \omega_{5}$ & $ \omega_{6}$ & $ h_{31}$ & $  h_{32}$ & $f_{31}$ & $f_{32}$ & $f_{33}$ & $ f_{5}$ & $  f_{7}$  \\ 
\noalign{\hrule height 1pt}
     $ \textbf{vac~1} $ & \multirow{3}{*}{$\textrm{Mkw}_{3}$} & $\mathcal{N}=0,4$ & $ \kappa $ & $ \xi$ & $  0 $ & $ 0 $ & $ 0 $ & $ 0 $ & $ 0 $ & $  0 $ & $0$ & $ \kappa$ & $ -\xi$ & $ 0 $ & $  0 $  \\ 
     \cline{1-1}\cline{3-16} 
     $ \textbf{vac~2}$ &  & $\mathcal{N}=0$ &  $ 0 $ & $ \kappa$ & $  0 $ & $0$ & $ 0$ & $ 0$ & $ 0$ & $  0$ & $0$ & $ 0$ & $ -\kappa$ & $ 0$ & $  0$  \\ 
     \cline{1-1}\cline{3-16}
     $ \textbf{vac~3}$ &  & $\mathcal{N}=0$ & $ 0$ & $ \kappa $ & $  \kappa$ & $0$ & $ 0$ & $ 0$ & $ 0$ & $  0$ & $0$ & $ 0$ & $ 0$ & $ 0$ & $  0$  \\ 
\noalign{\hrule height 1pt}
     $ \textbf{vac~4}$ & \multirow{2}{*}{${\textrm{AdS}_{3}}$} & $\mathcal{N}=4$ & $ 0$ & $ 0$ & $  0$ & $0$ & $ 0$ & $ \kappa$ & $ 0$ & $  0$ & $\pm\kappa$ & $ 0$ & $ 0$ & $ 0$ & $  -\kappa$ 
     \\ 
     \cline{1-1}\cline{3-16}
     $ \textbf{vac~5}$ &  & $\mathcal{N}=0$ & $ 0$ & $ 0$ & $  0$ & $0$ & $ 0$ & $ \kappa$ & $ 0$ & $  0$ & $\pm\kappa$ & $ 0$ & $ 0$ & $ 0$ & $  \kappa$ 
     \\
\noalign{\hrule height 1pt}
     $ \textbf{vac~6}$ & \multirow{2}{*}{$\textrm{AdS}_{3}$} & $\mathcal{N}=3$ & $ \kappa$ & $ 0$ & $  0$ & $0$ & $ -\kappa$ & $ \kappa$ & $ 0$ & $  0$ & $\pm\kappa $ & $ 0$ & $ \mp\kappa$ & $ 0$ & $  -2\kappa$  \\ 
     \cline{1-1}\cline{3-16}
     $ \textbf{vac~7}$ &  & $\mathcal{N}=1$ & $ \kappa$ & $ 0$ & $  0$ & $0$ & $ -\kappa$ & $ \kappa$ & $ 0$ & $  0$ & $\mp\kappa $ & $ 0$ & $ \pm\kappa$ & $ 0$ & $  2\kappa$  \\
\noalign{\hrule height 1pt}
     $ \textbf{vac~8}$ & \multirow{2}{*}{${\textrm{AdS}_{3}}$} & $\mathcal{N}=1$ & $ 0$ & $ 0$ & $  0$ & $0$ & $ -\kappa$ & $ \kappa$ & $ 0$ & $  0$ & $\pm\kappa$ & $ 0$ & $ 0$ & $ 0$ & $  \kappa$  \\ 
     \cline{1-1}\cline{3-16}
     $ \textbf{vac~9}$ &  & $\mathcal{N}=0$ & $ 0$ & $ 0$ & $  0$ & $0$ & $ -\kappa$ & $ \kappa$ & $ 0$ & $  0$ & $\pm\kappa$ & $ 0$ & $ 0$ & $ 0$ & $ - \kappa$  \\ 
\noalign{\hrule height 1pt}
  \cellcolor[HTML]{A3C1AD}$\textbf{vac~10}$ & \cellcolor[HTML]{A3C1AD}$\textrm{AdS}_{3}$ & \cellcolor[HTML]{A3C1AD}$\mathcal{N}=0$ & \cellcolor[HTML]{A3C1AD}$ 0$ & \cellcolor[HTML]{A3C1AD}$ 2\kappa$ & \cellcolor[HTML]{A3C1AD}$  \kappa$ & \cellcolor[HTML]{A3C1AD}$0$ & \cellcolor[HTML]{A3C1AD}$ 0$ & \cellcolor[HTML]{A3C1AD}$ 0$ & \cellcolor[HTML]{A3C1AD}$ 0$ & \cellcolor[HTML]{A3C1AD}$ 0$ & \cellcolor[HTML]{A3C1AD}$\kappa$ & \cellcolor[HTML]{A3C1AD}$ \pm\kappa$ & \cellcolor[HTML]{A3C1AD}$ -\kappa$ & \cellcolor[HTML]{A3C1AD}$ 0$ & \cellcolor[HTML]{A3C1AD}$  \pm\kappa$  \\
     \hline 
     \cellcolor[HTML]{A3C1AD}$\textbf{vac~11}$ & \cellcolor[HTML]{A3C1AD}$\textrm{AdS}_{3}$ & \cellcolor[HTML]{A3C1AD}$\mathcal{N}=0$ & \cellcolor[HTML]{A3C1AD}$ 0$ & \cellcolor[HTML]{A3C1AD}$ 2\kappa$ & $  \cellcolor[HTML]{A3C1AD}\kappa$ & \cellcolor[HTML]{A3C1AD}$0$ & \cellcolor[HTML]{A3C1AD}$ 0$ & \cellcolor[HTML]{A3C1AD}$ 0$ & \cellcolor[HTML]{A3C1AD}$ 0$ & \cellcolor[HTML]{A3C1AD}$  0$ & \cellcolor[HTML]{A3C1AD}$\kappa$ & \cellcolor[HTML]{A3C1AD}$ \pm\kappa$ & \cellcolor[HTML]{A3C1AD}$ -\kappa$ & \cellcolor[HTML]{A3C1AD}$ 0$ & \cellcolor[HTML]{A3C1AD}$  \mp\kappa$  \\
\noalign{\hrule height 1pt} 
     $ \textbf{vac~12}$ & \multirow{4}{*}{${\textrm{AdS}_{3}}$} & $\mathcal{N}=4$ & $ 0$ & $ 0$ & $  \mp\kappa$ & $\mp\kappa$ & $ \kappa$ & $ -2\kappa$ & $ 0$ & $  0$ & $\mp 2\kappa$ & $ 0$ & $ 0$ & $ 0$ & $  2\kappa$  \\
      \cline{1-1}\cline{3-16}
     $ \textbf{vac~13}$ &  & $\mathcal{N}=1$ & $ 0$ & $ 0$ & $  \mp\kappa$ & $\mp\kappa$ & $ \kappa$ & $ -2\kappa$ & $ 0$ & $  0$ & $\mp 2\kappa$ & $ 0$ & $ 0$ & $ 0$ & $  -2\kappa$  \\
     \cline{1-1}\cline{3-16}
     $ \textbf{vac~14}$ &  & $\mathcal{N}=0$ & $ 0$ & $ 0$ & $  \pm\kappa$ & $\pm\kappa$ & $ \kappa$ & $ -2\kappa$ & $ 0$ & $  0$ & $\mp 2\kappa$ & $ 0$ & $ 0$ & $ 0$ & $  2\kappa$  \\
     \cline{1-1}\cline{3-16}
     $ \textbf{vac~15}$ &  & $\mathcal{N}=0$ & $ 0$ & $ 0$ & $  \pm\kappa$ & $\pm\kappa$ & $ \kappa$ & $ -2\kappa$ & $ 0$ & $  0$ & $\mp 2\kappa$ & $ 0$ & $ 0$ & $ 0$ & $  -2\kappa$ \\
\noalign{\hrule height 1.5pt}
\end{tabular}}
\caption{Fluxes producing a type IIB with O$5$/D$5$ vacuum at the origin of moduli space. The \textbf{vac~1} is Mkw$_{3}$ and generically non-supersymmetric, but becomes $\mathcal{N}=4$ when $\kappa = \pm \xi$. For the AdS$_{3}$ supersymmetric vacua, the $\,\mathcal{N}=p\,$ supersymmetry is realised as $\,\mathcal{N}=(p,0)\,$ or $\,\mathcal{N}=(0,p)\,$  depending on the upper/lower sign choice of the fluxes. \textbf{Vac~10} and \textbf{vac~11} have been highlighted for future reference in the main text.}
\label{Table:flux_vacua_IIB}
\end{center}
\end{table}

A quick inspection of Table~\ref{Table:flux_vacua_IIB} shows that there are both Mkw$_{3}$ as well as  AdS$_{3}$ flux vacua. Both types of vacua can be either supersymmetric or non-supersymmetric depending on specific sign choices of the flux parameters. Very importantly, all the vacua are compatible with
\begin{equation}
H_{(3)} = F_{(5)} = 0 \ .
\end{equation} 
This fact, together with the vanishing of the axionic components of the complex fields (their real part in our parameterisation), guarantees that \textit{all} the vacua in Table~\ref{Table:flux_vacua_IIB} can be reinterpreted within the context of $\,\mathcal{N}=1\,$ co-calibrated G$_{2}$ orientifold compactifications of \cite{Emelin:2021gzx}. We have verified that, indeed,  all the vacua in Table~\ref{Table:flux_vacua_IIB} are also extrema of the scalar potential of \cite{Emelin:2021gzx} upon a straightforward mapping of flux parameters and moduli fields. Using the correspondence between our three-dimensional setup and the top-down construction of \cite{Emelin:2021gzx}, the (string frame) volume of the seven-dimensional internal space, $\textrm{vol}_{7}$, and the string coupling constant, $g_{s}$, are given by
\begin{equation}
\label{vol7&gs_IIB}
\left(\textrm{vol}_{7}\right)^{4} = \frac{(\textrm{Im}T)^{3}}{(\textrm{Im}R)^{2} \, (\textrm{Im}S) \, (\textrm{Im}U)^{12}} 
\hspace{10mm} \textrm{ and } \hspace{10mm}
g_{s}^2 = \frac{1}{(\textrm{Im}U)^{3} \, \textrm{Im}R}  \  ,
\end{equation}
in terms of (the imaginary parts of) our moduli fields. Since all the vacua in Table~\ref{Table:flux_vacua_IIB} are at the origin of moduli space, see (\ref{GTTO_1})-(\ref{GTTO_2}), one automatically gets $\textrm{vol}_{7}=g_{s}=1$ for all of them. However, as previously explained, we can move the moduli fields in (\ref{vol7&gs_IIB}) away from the origin of moduli space (therefore changing $\textrm{vol}_{7}$ and $g_{s}$) just by rescaling the flux parameters. Therefore, in order to study some phenomenological aspects of the AdS$_{3}$ vacua we found, in particular, scale separation between AdS$_{3}$ and the internal space, it will be convenient to resort to a more standard picture in which the moduli VEV's are expressed in terms of the flux parameters. Once this is done, it turns out that only \textbf{vac~10} and \textbf{vac~11} in Table~\ref{Table:flux_vacua_IIB} (which are highlighted therein) stabilise the four imaginary parts of the complex scalars and yield well-defined $\textrm{vol}_{7}$ and $g_{s}$ in (\ref{vol7&gs_IIB}).

\begin{table}[t]
\begin{center}
\scalebox{0.8}{
\renewcommand{\arraystretch}{1.5}
\begin{tabular}{!{\vrule width 1.5pt}c!{\vrule width 1pt}c!{\vrule width 1pt}c!{\vrule width 1pt}cccccc!{\vrule width 1pt}cc!{\vrule width 1pt}ccc!{\vrule width 1pt}c!{\vrule width 1pt}c!{\vrule width 1.5pt}}
\noalign{\hrule height 1.5pt}
     ID & Scalar spectrum  \\ 
\noalign{\hrule height 1pt}
     $ \textbf{vac~1} $ & $ g^{-2} \, m^2 =  0_{(30)}, \left(\frac{\kappa^2}{16} \right)_{(9)}, \left(\frac{\kappa^2}4\right)_{(9)} ,
     \left(\frac{\xi^2}4\right)_{(9)}  , 
     \left(\frac{9\kappa^2}{16} \right)_{(1)}, \left[ \frac{\left( \kappa - 2\xi \right)^2}{16} \right]_{(3)}, \left[ \frac{\left( \kappa + 2\xi \right)^2}{16} \right]_{(3)}  $  \\[2mm] 
     \hline 
     $ \textbf{vac~2} $ & \multirow{2}{*}{$g^{-2} \, m^2 =  \left( \frac{\kappa^2}{4} \right)_{(15)} , 0_{(49)} $}  \\ 
     \cline{1-1} 
     $ \textbf{vac~3} $ &  \\ 
\noalign{\hrule height 1pt}
     $ \textbf{vac~4} $ & \multirow{2}{*}{$\begin{array}{ccl}
        m^2 L^2 &=&  8_{(19)}, \ 0_{(45)}  \\[-2mm]
         \Delta &=&  4_{(19)}, \ 2_{(45)} 
     \end{array}$}
     \\ 
    \cline{1-1} 
     $ \textbf{vac~5} $ &   
     \\
\noalign{\hrule height 1pt}
     $ \textbf{vac~6} $ & \multirow{2}{*}{$\begin{array}{ccl}
     m^2 L^2 &=&  8_{(10)}, \ 4_{(18)} , \ 0_{(36)}  \\[-2mm]
      \Delta &=&  4_{(10)}, \ (1+\sqrt{5})_{(18)} , \ 2_{(36)}
     \end{array}$}  \\ 
     \cline{1-1} 
     $ \textbf{vac~7} $ &   \\
\noalign{\hrule height 1pt}
     $ \textbf{vac~8} $ & \multirow{2}{*}{$\begin{array}{ccl}
     m^2 L^2 &=& 24_{(10)}, \ 8_{(25)}, \ 0_{(29)} \\[-2mm]
      \Delta &=& 6_{(10)}, \ 4_{(25)}, \ 2_{(29)} \end{array}$} \\ 
     \cline{1-1} 
     $ \textbf{vac~9} $ &   \\ 
\noalign{\hrule height 1pt}
      \cellcolor[HTML]{A3C1AD}$ \textbf{vac~10} $ &  \cellcolor[HTML]{A3C1AD}$\begin{array}{ccl}
     m^2 L^2 &=&  80_{(3)}, \ 48_{(9)}, \ 24_{(4)}, \ 8_{(7)}, \ 0_{(41)} \\[-2mm]
     \Delta &=&  10_{(3)}, \ 8_{(9)}, \ 6_{(4)}, \ 4_{(7)}, \ 2_{(41)}
     \end{array}$   \\[1mm]
     \hline 
     \cellcolor[HTML]{A3C1AD}$ \textbf{vac~11} $ &  \cellcolor[HTML]{A3C1AD}$\begin{array}{ccl} 
     m^2 L^2 &=& 48_{(15)}, \ 8_{(13)}, \ 0_{(36)} \\[-2mm]
     \Delta &=& 8_{(15)}, \ 4_{(13)}, \ 2_{(36)}
     \end{array}$   \\[1mm]
\noalign{\hrule height 1pt} 
     $ \textbf{vac~12} $ & \multirow{4}{*}{$\begin{array}{ccl}
     m^2 L^2 &=&  15_{(8)}, \ 8_{(19)}, \ 3_{(8)}, \ 0_{(29)} \\[-2mm]
     \Delta &=&  5_{(8)}, \ 4_{(19)}, \ 3_{(8)}, \ 2_{(29)}
     \end{array}$}  \\
     \cline{1-1} 
     $ \textbf{vac~13} $ &    \\
     \cline{1-1}
     $ \textbf{vac~14} $ &    \\
     \cline{1-1}
     $ \textbf{vac~15} $ &  \\
\noalign{\hrule height 1.5pt}
\end{tabular}}
\caption{The $64$ scalar masses at the type~IIB with O$5$/D$5$ vacua of Table~\ref{Table:flux_vacua_IIB}. The subscript in $n_{(s)}$ denotes the multiplicity of the mass $n$ in the scalar spectrum. \textbf{Vac~10} and \textbf{vac~11} have been highlighted for future reference in the main text.}
\label{Table:flux_vacua_IIB_masses}
\end{center}
\end{table}

As far as stability is concerned, it is worth highlighting that negative masses turn out to be always absent in the spectrum of scalar fluctuations, thus making all the vacua in Table~\ref{Table:flux_vacua_IIB} perturbatively stable within half-maximal supergravity. We have collected the $64$ scalar masses at each vacuum in Table~\ref{Table:flux_vacua_IIB_masses}. For the AdS$_{3}$ vacua, we have normalised the scalar spectrum using the AdS$_{3}$ radius, which is $\,L^2=-2/V_{0}\,$ in 3D Planck units. The 3D Planck constant (in string units) is defined as $\,m_P = g_s^{-2} \, \textrm{vol}_7\,$ \cite{Farakos:2020phe}. We have also included the conformal dimension $\,\Delta\,$ of the would-be dual  CFT$_{2}$ operators. More concretely, $\,\Delta\,$ corresponds to the larger root of
\begin{equation}
\label{Delta_d=2}
m^2 L^2 = \Delta (\Delta - d) \hspace{8mm} \textrm{ with } \hspace{8mm} d=2 \ .
\end{equation}

\subsubsection{Integer $\Delta$'s and polynomial shift symmetries}

It was beautifully observed in \cite{Apers:2022vfp} that a polynomial shift symmetry mechanism is at play in the DGKT type IIA compactifications \cite{DeWolfe:2005uu}, accounting for the emergence of integer-valued conformal dimensions $\Delta$'s. More concretely, a scalar field $\,\phi\,$ in AdS$_{d+1}$ enjoys a \textit{level}~$k\,$ polynomial shift symmetry of the form
\begin{equation}
\phi \rightarrow \phi + c_{\mu_{1} \cdots \mu_{k}} \, X^{\mu_{1}} \cdots X^{\mu_{k}} \ ,
\end{equation}
where $\,c_{\mu_{1} \cdots \mu_{k}}\,$ is a rank-$k$ symmetric traceless constant tensor and $X^{\mu}$ are coordinates on an embedding $(d+2)$-dimensional flat spacetime, if its mass is given by 
\begin{equation}
m^{2} L^{2} = k (k+d) \ ,
\end{equation}
so that $\,\Delta_{-} = -k\,$ (smaller root) and $\,\Delta_{+}=d+k\,$ (larger root) \cite{Bonifacio:2018zex}. For a scalar field in AdS$_{3}$, a level $k$ polynomial shift symmetry requires
\begin{equation}
m^{2} L^{2} = k (k+2) 
\hspace{5mm} \Rightarrow \hspace{5mm}
\Delta_{-} = -k \,\,\,\, \textrm{ and } \,\,\,\, \Delta_{+}=2+k \ .
\end{equation}
and the dual operator has conformal dimension $\,\Delta=2+k$. One then finds values of the form
\begin{equation}
\label{symmetric_values}
\begin{array}{rcccccccccccccccccccc}
k &=& 0 & , & 1 & , & 2 & , & 3  & , & 4 & , & 5 & , & 6 & , & 7 & , & 8 & , & \ldots \\[2mm]
m^2 L^2 &=&  0 & , & 3 & , & 8 & , & 15 & , & 24  & , & 35 & , & 48 & , & 63 & , & 80  & , & \ldots \\[2mm]
\Delta &=&  2 & , & 3 & , & 4 & , & 5  & , & 6 & , & 7 & , & 8 & , & 9 & , & 10 & , & \ldots
\end{array} 
\end{equation}

Remarkably, and with the exception of \textbf{vac~6} and \textbf{vac~7} which contain a mass $\,m^{2}L^{2}=4$, all the AdS$_{3}$ vacua in Table~\ref{Table:flux_vacua_IIB_masses} come along with normalised scalar masses and $\Delta$'s contained in (\ref{symmetric_values}), thus pointing at a polynomial shift symmetry mechanism, as originally observed for DGKT vacua in \cite{Apers:2022vfp}. It would be interesting to further explore this observation and to identify the microscopic origin, if any, of this polynomial shift symmetry. It could just be a bonus symmetry associated to the simplicity of our compactification scheme: we are just allowing for the single type of O$5$/D$5$ sources in (\ref{O5_location}). This is to be contrasted with more complex models involving intersecting O$p$/D$p$ sources, which have been explored in the literature and typically feature non-integer $\Delta$'s \cite{Emelin:2021gzx,Apers:2022vfp,Apers:2022zjx}.\footnote{See ref.~\cite{Farakos:2025bwf} for a recent example of a type IIA compactification with intersecting O$6$-planes still yielding integer-valued $\Delta$'s.}

\subsubsection{AdS$_{3}$ flux vacua with scale separation, anisotropy and integer $\Delta$'s}
\label{sec:scale_sep&int_Deltas_in_vac10,11}

Let us focus for the rest of the section on \textbf{vac~10} and \textbf{vac~11} in Tables \ref{Table:flux_vacua_IIB} and \ref{Table:flux_vacua_IIB_masses}. Moving them away from the origin of moduli space via flux rescalings, and expressing the new VEV's in terms of the fluxes one obtains
\begin{equation}
\label{vac10/11_VEVs}
\textrm{Im}R =  -\frac{f_{31} \left( \pm f_{32}^3 \, f_7\right)^{\frac12}}{\omega_3^2 \, f_{33}}
\hspace{2mm} , \hspace{2mm} 
\textrm{Im}S = \frac{\omega_3 \, f_{33}^2}{f_{31} \left( \pm f_{32}^3 \, f_7\right)^{\frac12}} 
\hspace{2mm} , \hspace{2mm} 
\textrm{Im}T = -\frac{\left( \pm f_{32}^3 \, f_7\right)^{\frac12}}{\omega_3 \, f_{33}}
\hspace{2mm} , \hspace{2mm} 
\textrm{Im}U = \left( \pm \frac{f_{32}}{f_7} \right)^{\frac12} \ ,
\end{equation}
where the upper (lower) sign choice corresponds to \textbf{vac~10} (\textbf{vac~11}). Positivity of the scalar VEV's in (\ref{vac10/11_VEVs}) then imposes
\begin{equation}
\label{vac_signs}
\pm f_{32} \, f_{7} > 0
\hspace{6mm} , \hspace{6mm} 
\omega_{3} \, f_{31} >0
\hspace{6mm} , \hspace{6mm} 
\omega_{3} \, f_{33} < 0
\hspace{6mm} \textrm{ and } \hspace{6mm}  
f_{31} \, f_{33}  < 0 \ .
\end{equation}
Additionally, there is the relation
\begin{equation}
\label{vac10/11_additional_relations}
\omega_2 \, f_{31} + 2 \, \omega_3 \, f_{33} = 0  \quad  \Rightarrow \quad    \omega_2 =  - 2 \, \omega_3 \, \frac{f_{33}}{ f_{31}}   \ ,
\end{equation}
that arises as a tadpole cancellation condition for O$5$/D$5$ sources different from the ones in (\ref{O5_location}). In other words, eq.~(\ref{vac10/11_additional_relations}) is the only non-trivial condition arising from the right-hand-side equation in (\ref{Tadpole_IIB_O5}). Any other flux parameter in the RSTU-model not entering (\ref{vac10/11_VEVs})-(\ref{vac10/11_additional_relations}) must  vanish identically at \textbf{vac~10} and \textbf{vac~11}. Summarising, the set of non-zero flux parameters consists of
\begin{equation}
\left\lbrace f_{7} \,,\, f_{31} \, ,\, f_{32} \, ,\, f_{33} \,,\,  \omega_{2}  \, ,\,  \omega_{3}  \right\rbrace
\end{equation}
subject to the tadpole cancellation condition (\ref{vac10/11_additional_relations}). Very importantly, the unrestricted flux-induced tadpole in (\ref{Tadpole_O5/D5_allowed}) for the allowed O$5$/D$5$ sources in (\ref{O5_location}) translates into
\begin{equation}
\label{tadpole_vac_10&11}
6\,  (\omega_3 f_{33}) \, \left(\frac{f_{33}}{f_{31}}\right)  = J_{\textrm{O}5/\textrm{D}5} \equiv J_{\textrm{O}5} -  J_{\textrm{D}5} >0 \ .
\end{equation}
Therefore, both \textbf{vac~10} and \textbf{vac~11} require an excess of O$5$-planes over the number of D$5$-branes. Since the number of O$5$-planes has an upper bound given by the number of fixed points of the orientifold involution, the left-hand-side in (\ref{tadpole_vac_10&11}) is also bounded from above. Lastly, a direct evaluation of the scalar potential at \textbf{vac~10} and \textbf{vac~11}, as well as the internal volume $\textrm{vol}_{7}$ and the string coupling $g_{s}$ in (\ref{vol7&gs_IIB}), yields
\begin{equation}
\label{L&vol7_vac_10,11}
L^2 = 64 \frac{(f_{32}^3 f_7)^2}{(\omega_3 f_{33})^6} \, f_{31}^2
\hspace{5mm} , \hspace{5mm}  
(\textrm{vol}_{7})^{4} =  \mp  \frac{f_7^7}{f_{31} f_{32}^3 f_{33}^3} 
\hspace{5mm} , \hspace{5mm}
g_s^{2} = \mp \frac{(\omega_3 f_{33}) f_{7}^{2} \, \omega_{3}^{2}}{(\omega_3 f_{31})(f_{32}^3 f_7)} \ ,
\end{equation}
where $\,L^2 = - 2/V_{0}\,$ denotes the AdS$_3$ radius in 3D Planck units.

\subsubsection*{Scale separation}

As a proof of concepts, we will show an example that achieves scale separation, $m_P^{-1} L \gg \left(\textrm{vol}_{7}\right)^{\frac{1}{7}}$,\footnote{Note the factor $\,m_P^{-1}\,$ which is necessary to express $\,L\,$ in string units.} and weak coupling, $\,g_{s} \ll 1$, in a parametrically-controlled manner while keeping $\,J_{\textrm{O}5/\textrm{D}5} =\textrm{fixed}$. Let us first parameterise the fluxes as\footnote{The example presented here is distinct from that analysed in ref.~\cite{Arboleya:2024vnp}.}
\begin{equation}
\label{flux_example}
f_{7}= 2 \, \alpha^{2} \, \beta \, N^{11}
\hspace{3mm} , \hspace{3mm}
f_{31} = 2 \, \alpha \, \beta^{2}
\hspace{3mm} , \hspace{3mm}
f_{32} = \alpha \, \beta \, N^{5}
\hspace{3mm} , \hspace{3mm}
f_{33} = - \alpha^{2}
\hspace{3mm} , \hspace{3mm}
\omega_{2} = \alpha
\hspace{3mm} , \hspace{3mm}
\omega_{3} = \beta^{2} \ ,
\end{equation}
with $\,\alpha,\beta,N \in \mathbb{N}\,$ so that the fluxes are integer-valued and the condition (\ref{vac10/11_additional_relations}) is automatically satisfied. The fluxes in (\ref{flux_example}) yield
\begin{equation}
\label{scales_example}
g_{s} = N^{-2}
\hspace{3mm} , \hspace{3mm}
\textrm{vol}_{7} = 2^{\frac{3}{2}} \, \alpha \, \beta^{\frac{1}{2}} \, N^{\frac{31}2}
\hspace{3mm} , \hspace{3mm}
m_P^{-1} L = 2^{\frac72} \alpha^{-1} \beta^{-\frac12} \,  N^{\frac{13}2}
\hspace{3mm} , \hspace{3mm}
J_{\textrm{O}5/\textrm{D}5} = 3 \, \alpha^{3} \ ,
\end{equation}
so the parameter $\,N\,$ plays the role of a scaling parameter which must be taken arbitrarily large (formally $N\rightarrow \infty$) in order to get a regime of parametrically-controlled scale separation and weak coupling without affecting $J_{\textrm{O}5/\textrm{D}5}$. Setting for example $\,\alpha=1$, then the compactification scheme will include the O5-planes located at the fixed points of the orientifold involution, but also a number of D5-branes sitting on top of them such that $\,J_{\textrm{O}5/\textrm{D}5} \equiv J_{\textrm{O}5}-J_{\textrm{D}5} = 3$.

\subsubsection*{Anisotropy}

Let us briefly touch upon the issue of anisotropy of the internal space versus scale separation. When expressed in the string frame -- we add a superscript $\,^{s}\,$ -- all the internal cycles in our example have a parametrically large volume. In particular, the characteristic sizes of the would-be one-cycles scale as
\begin{equation}
\label{scalings_one-fomrs_example}
L^{s}_{a}\,,\,L^{s}_{i}\sim N^{\frac{3}{2}}
\hspace{8mm} \textrm{ and } \hspace{8mm}
L^{s}_{7}\sim N^{\frac{13}{2}}  \ .
\end{equation}
Consequently: three-cycles scale as $\,L^{s}_{abk},\,L^{s}_{ijk}\sim N^{\frac{9}{2}}\,$ and $\,L^{s}_{ai7} \sim N^{\frac{19}{2}}$; four-cycles scale as $\,L^{s}_{ijc7},\,L^{s}_{abc7}\sim N^{\frac{22}{2}}\,$ and $\,L^{s}_{aibj} \sim N^{\frac{12}{2}}$; the internal volume scales as $\,\textrm{vol}_{7} \sim N^{\frac{31}{2}}$. It is worth highlighting the different scalings of the would-be one-cycles in (\ref{scalings_one-fomrs_example}), yielding an anisotropic internal space with the $7^{\textrm{th}}$ direction playing a distinct role. This is consistent with the obstruction to achieving parametric scale separation in isotropic compactifications recently put forward in \cite{Tringas:2025uyg}.

Let us take a closer look at the algebra spanned by the isometry generators $(X_{a},X_{i})$ and $\,X_{7}\,$ on the anisotropic internal space. For the metric fluxes in (\ref{flux_example}), the commutation relations in (\ref{X_commutators}) reduce to
\begin{equation}
\label{algebra_isometry_example}
\begin{array}{cccc}
\left[ X_{2}, X_{7} \right]  = \beta^2  \, X_{1} & \hspace{8mm} , & \hspace{8mm}  \left[ X_{1},X_{7} \right] = - \alpha \, X_{2}  & , \\[2mm] 
\left[ X_{4}, X_{7} \right]  = \beta^2 \, X_{3} & \hspace{8mm} , & \hspace{8mm}  \left[ X_{3},X_{7} \right] = - \alpha \, X_{4}  & , \\[2mm] 
\left[ X_{6}, X_{7} \right]  = \beta^2 \, X_{5} & \hspace{8mm} , & \hspace{8mm}  \left[ X_{5},X_{7} \right] = - \alpha \, X_{6}  & , 
\end{array}
\end{equation}
which correspond to a $2$-step solvable algebra with degenerate Killing--Cartan metric. The algebra (\ref{algebra_isometry_example}) describes a particular example of a seven-dimensional solvmanifold (see eq.$(2.2)$ of \cite{VanHemelryck:2025qok}) for which the Maurer--Cartan equation (\ref{structure_equation}) can be integrated. The resulting seven-dimensional solvmanifold can be seen as three copies of the three-dimensional solvmanifold E$_2$ sharing the seventh direction \cite{VanHemelryck:2025qok}.

\subsubsection*{Integer conformal dimensions}

As highlighted in Table~\ref{Table:flux_vacua_IIB_masses}, both \textbf{vac~10} and \textbf{vac~11} come along with a spectrum of scalar perturbations within half-maximal supergravity that contains only non-negative masses. In particular, the normalised mass spectra are given by
\begin{equation}
\label{spectrum_vac10}
\textbf{vac~10}: \hspace{5mm}
\begin{array}{ccl}
m^2 L^2 &=&  80_{(3)} \, , \, 48_{(9)} \, , \,  24_{(4)} \, , \, 8_{(7)} \, , \, 0_{(41)} \ , \\[2mm]
\Delta &=&  10_{(3)} \, , \, 8_{(9)} \, , \, 6_{(4)} \, , \, 4_{(7)} \, , \, 2_{(41)} \ , 
\end{array} 
\end{equation}
and
\begin{equation}
\label{spectrum_vac11}
\textbf{vac~11}: \hspace{5mm}
\begin{array}{ccl}
m^2 L^2 &=& 48_{(15)} \,,\,  8_{(13)} \,,\, 0_{(36)}  \ , \\[2mm]
\Delta &=& 8_{(15)} \,,\, 4_{(13)} \,,\, 2_{(36)} \ .
\end{array} 
\end{equation}
Therefore, although \textbf{vac~10} and \textbf{vac~11} are related by a flip of sign of the flux parameter $\,f_{7}\,$ in (\ref{vac10/11_VEVs}) (see also Table~\ref{Table:flux_vacua_IIB}), their mass spectra are different. Remarkably, and despite the absence of supersymmetry, all the $\,64\,$ conformal dimensions for the would-be dual CFT$_{2}$ operators  turn out to be integer-valued. The values of the $\Delta$'s in (\ref{spectrum_vac10}) and (\ref{spectrum_vac11}) are contained in (\ref{symmetric_values}), pointing at polynomial shift symmetries with $\,k=8,6,4,2,0\,$ and $\,k=6,2,0$, respectively.

\subsection{Beyond the RSTU-model: breaking $\,\textrm{SO}(3)\,$ invariance}
\label{sec:susy_cousin}

Let us close this section by discussing a supersymmetric cousin of our solutions \textbf{vac~10} and \textbf{vac~11} found in \cite{VanHemelryck:2025qok}. This supersymmetric solution is \textit{not} compatible with the SO(3) symmetry of the RSTU-models. For example, when plugging the metric fluxes of \cite{VanHemelryck:2025qok} into (\ref{X_commutators}), the isometry algebra in (\ref{algebra_isometry_example}) gets modified to
\begin{equation}
\label{algebra_isometry_susy}
\begin{array}{cccc}
\left[ X_{2}, X_{7} \right]  = \mp \beta^2  \, X_{1} & \hspace{8mm} , & \hspace{8mm}  \left[ X_{1},X_{7} \right] = \pm \alpha \, X_{2}  & , \\[2mm] 
\left[ X_{4}, X_{7} \right]  = \pm \beta^2 \, X_{3} & \hspace{8mm} , & \hspace{8mm}  \left[ X_{3},X_{7} \right] = \mp \alpha \, X_{4}  & , \\[2mm] 
\left[ X_{6}, X_{7} \right]  = \pm  \beta^2 \, X_{5} & \hspace{8mm} , & \hspace{8mm}  \left[ X_{5},X_{7} \right] = \mp \alpha \, X_{6}  & ,
\end{array}
\end{equation}
and can no longer be solely written in terms of the two $\textrm{SO}(3)$-invariant flux parameters $\,\omega_{2}\,$ and $\,\omega_{3}\,$ in (\ref{metric_flux_SO(3)}).

We present in these proceedings the full spectrum of scalar fluctuations within half-maximal supergravity for the supersymmetric cousin of \textbf{vac10} and \textbf{vac11} found in \cite{VanHemelryck:2025qok}. It is given by
\begin{equation}
\label{spectrum_susy}
\textbf{SUSY vacuum} : \hspace{5mm}
\begin{array}{ccl}
m^2 L^2 &=&  80_{(4)} \, , \, 48_{(7)} \, , \,  24_{(4)} \, , \, 8_{(9)} \, , \, 0_{(40)} \ , \\[2mm]
\Delta &=&  10_{(4)} \, , \, 8_{(7)} \, , \, 6_{(4)} \, , \, 4_{(9)} \, , \, 2_{(40)} \ .
\end{array} 
\end{equation}
As a result, the values of the normalised masses (and corresponding $\Delta$'s) are the same as in the non-supersymmetric \textbf{vac~10} in (\ref{spectrum_vac10}), but the multiplicities are slightly modified. We have also verified that the supersymmetric vacuum preserves $\,\mathcal{N}=(1,0)\,$ or $\,\mathcal{N}=(0,1)\,$ supersymmetry within half-maximal supergravity depending on the upper/lower choice of sign in (\ref{algebra_isometry_susy}). The normalised masses and $\Delta$'s in (\ref{spectrum_susy}) belong to the list in (\ref{symmetric_values}), once again pointing at polynomial shift symmetries with $\,k=8,6,4,2,0$.

\section{Summary, open questions and future directions}

In these proceedings, we have reviewed recent progress in the search for type II orientifold reductions that accommodate AdS$_{3}$ flux vacua with various key features: small string coupling, large internal volume, parametrically-controlled scale separation between AdS$_{3}$ and the internal space, quantised fluxes, and integer-valued conformal dimensions for the would-be dual CFT$_{2}$ operators. The first step in this direction was taken in \cite{Farakos:2020phe} where the authors investigated type IIA reductions on G$_2$-holonomy spaces with O$2$/O$6$-planes. While most of the key features are achieved in the AdS$_{3}$ flux vacua of \cite{Farakos:2020phe}, the conformal dimensions of the would-be dual operators are not integer-valued \cite{Apers:2022zjx} (see ref.~\cite{Farakos:2025bwf} for some recent examples with integer $\Delta$'s).  It was later observed in \cite{Apers:2022zjx} that, unlike for the DGKT vacua \cite{DeWolfe:2005uu}, the AdS$_{3}$ flux vacua constructed in \cite{Farakos:2020phe} are incompatible with a refined version \cite{Buratti:2020kda} of the AdS distance conjecture \cite{Lust:2019zwm}, owing to the absence of the necessary discrete higher-form symmetries. These arguments placed the type IIA AdS$_{3}$ flux vacua of \cite{Farakos:2020phe} in the Swampland. 

It turns then natural to resort to the type IIB reductions on co-calibrated G$_{2}$ orientifolds with O9/O5-planes of \cite{Emelin:2021gzx}, which also accommodate AdS$_{3}$ flux vacua. Unfortunately, it was first argued in \cite{Emelin:2021gzx} that scale separation seems to be hindered by the quantisation of the geometric fluxes, and then observed in \cite{Apers:2022zjx,Apers:2022vfp} that the $\Delta$'s of the dual operators are not integer. In the midst of this bleak outlook, ref.~\cite{Arboleya:2024vnp} took a second look at a simple class of type IIB orientifold reductions with just a single type of O5-planes (and D5-branes), and showed that (classical) AdS$_{3}$ flux vacua with all the key features were indeed possible. Namely, type IIB AdS$_{3}$ flux vacua with small string coupling, large internal volume, parametrically-controlled scale separation, quantised fluxes, and integer-valued conformal dimensions for the would-be dual CFT$_{2}$ operators. The AdS$_{3}$ flux vacua presented in \cite{Arboleya:2024vnp}, along with their supersymmetric counterpart identified in \cite{VanHemelryck:2025qok}, are the solutions we have re-examined in these proceedings.

A list of open questions and future directions to explore includes:
\begin{itemize}

\item Building on the analysis of \cite{Apers:2022zjx}, it would be worthwhile to investigate whether the discrete higher-form symmetries required by \cite{Buratti:2020kda} -- which rule in the DGKT vacua \cite{DeWolfe:2005uu} and rule out the type IIA AdS$_{3}$ flux vacua of \cite{Farakos:2020phe} -- are present/absent in the type IIB AdS$_{3}$ flux vacua of \cite{Arboleya:2024vnp,VanHemelryck:2025qok}.

\item The simplicity of our compactification scheme -- which just contains a single type of O$5$/D$5$ sources -- allows for an exploration of various aspects regarding the sources. The first one is to investigate whether the smeared limit of O$5$/D$5$ sources that we have considered here provides a valid approximation to full-fledged AdS$_3$ solutions in string theory. In other words, it would be interesting to analyse backreaction issues order by order in $\,g_{s}\,$ along the lines of \cite{Junghans:2020acz, Junghans:2023yue, Emelin:2022cac, Emelin:2024vug}. The second direction is to investigate the dynamics of open string modes in our RSTU-models with O$5$/D$5$ sources \cite{Arboleya:2025lwu}. More concretely, the half-maximal supersymmetry of the three-dimensional models opens up the possibility for a search of potential instabilities of the non-supersymmetric vacua within the open string sector, as recently done in \cite{Balaguer:2024cyb} for the half-maximal $D=4$ supergravities of \cite{Dibitetto:2011gm} arising from type IIA orientifold reductions in the presence of a single type of O$6$/D$6$ sources.

\item It would be interesting to carry out a detailed analysis of the Kaluza--Klein (KK) mass spectrum at the non-supersymmetric AdS$_{3}$ vacua reviewed in these proceedings for two main reasons. Firstly, to study their higher-dimensional stability. Secondly, in the special case of \textbf{vac~10} and \textbf{vac~11}, to establish whether the KK modes are physically decoupled from the degrees of freedom of the three-dimensional supergravity despite the characteristic size of the would-be one-cycle $\,L_7^s\,$ in (\ref{scalings_one-fomrs_example}) becomes comparable to the AdS$_{3}$ radius in (\ref{scales_example}), as originally noticed in \cite{VanHemelryck:2025qok}.\footnote{We thank Vincent Van Hemelryck for pointing this out to us.} The tower of KK modes could be accessible via exceptional field theory techniques \cite{Malek:2019eaz} (see \cite{Eloy:2020uix,Eloy:2023acy,Eloy:2024lwn} for some examples of AdS$_{3}$ KK spectrometry), suitably adapted to the context of type II orientifold compactifications. This could shed some new light on how to make sense of scale separation and positive masses from a CFT viewpoint \cite{Gautason:2018gln,Collins:2022nux}.

\item It would also be interesting to explore the landscape of vacua -- both within RSTU models and beyond -- for other possible type II duality frames involving a single type of O$p$/D$p$ sources with $p \neq 2,5$ (see \cite{Arboleya:2024vnp} for a detailed discussion). These configurations likewise give rise to half-maximal supergravities in $\,D=3$. Such an investigation could provide valuable insight into how prevalent scale-separated AdS$_3$ flux vacua are within type II string theory \cite{wip_landscape}.

\end{itemize}

Finally, according to \cite{Ooguri:2016pdq}, non-supersymmetric anti-de Sitter vacua supported by fluxes must be unstable. Alternatively, there are no non-supersymmetric conformal field theories whose holographic duals have a description in terms of Einstein gravity coupled to a finite number of matter fields, precisely as our three-dimensional half-maximal supergravities. However, being in three dimensions where vectors are dualised into scalars could weaken the range of applicability of the various arguments put forward in the context of the Weak Gravity Conjecture \cite{Arkani-Hamed:2006emk} and its various subsequent sharpenings. Also, the fact that there is a supersymmetric (therefore stable) version \cite{VanHemelryck:2025qok} of the non-supersymmetric, scale-separated AdS$_{3}$ flux vacua with integer $\Delta$'s found in \cite{Arboleya:2024vnp} motivates further investigation on the stability issue. We hope to come back to this, along with the open questions outlined above, in the near future.

\section*{Acknowledgements}

We are grateful to Giuseppe Sudano, Thomas Van Riet and Vincent Van Hemelryck for interesting discussions. This work is supported by the Spanish national grant MCIU-22-PID2021-123021NB-I00.

\bibliographystyle{JHEP}
\bibliography{references}

\end{document}